\documentclass[twocolumn]{aastex63}
\usepackage[utf8]{inputenc}

\received{January 20, 2021}
\accepted{March 25, 2021}

\submitjournal{Astrophysical Journal (ApJ)}
\shorttitle{Coronal Hole Boundaries and Automated Detection Schemes}
\shortauthors{Reiss et al.}

\usepackage{color}

\begin{document}
\title{The Observational Uncertainty of Coronal Hole Boundaries in Automated Detection Schemes}

\correspondingauthor{Martin A. Reiss}
\email{martin.reiss@oeaw.ac.at}

\author[0000-0002-6362-5054]{Martin A.~Reiss}
\affiliation{Space Research Institute, 
Austrian Academy of Sciences, 
Graz, Austria}

\author[0000-0002-5547-9683]{Karin Muglach}
\affiliation{Heliophysics Science Division, 
NASA Goddard Space Flight Center, 
Greenbelt, MD 20771, USA}
\affiliation{Catholic University of America, Washington, DC 20064, USA}

\author[0000-0001-6868-4152]{Christian M\"ostl}
\affiliation{Space Research Institute, 
Austrian Academy of Sciences, 
Graz, Austria}

\author{Charles N.~Arge}
\affiliation{Heliophysics Science Division, 
NASA Goddard Space Flight Center, 
Greenbelt, MD 20771, USA}

\author[0000-0003-2021-6557]{Rachel Bailey}
\affiliation{Conrad Observatory, Zentralanstalt für Meteorologie und Geodynamik, Vienna, Austria}

\author{Véronique Delouille}
\affiliation{Royal Observatory of Belgium, Brussels, Belgium}

\author{Tadhg M. Garton}
\affiliation{University of Southampton, Southampton, UK}

\author{Amr Hamada}
\affiliation{Space Physics and Astronomy Research Unit, Space Climate Group, University of Oulu, Oulu, Finland}

\author{Stefan Hofmeister}
\affiliation{University of Graz, Institute of Physics, Graz, Austria}
\affiliation{Columbia Astrophysics Laboratory, Columbia University, New York, NY 10027, USA}

\author{Egor Illarionov}
\affiliation{Moscow State University, Moscow, 119991, Russia}
\affiliation{Moscow Center of Fundamental and Applied Mathematics, Moscow, 119234, Russia}

\author{Robert Jarolim}
\affiliation{University of Graz, Institute of Physics, Graz, Austria}

\author[0000-0001-9874-1429]{Michael S.F. Kirk}
\affiliation{Heliophysics Science Division, 
NASA Goddard Space Flight Center, 
Greenbelt, MD 20771, USA}
\affiliation{ASTRA llc, Louisville, CO, 80027, USA}

\author{Alexander Kosovichev}
\affiliation{Center for Computational Heliophysics, New Jersey Institute of Technology, Newark, NJ 07102, USA}
\affiliation{Department of Physics, New Jersey Institute of Technology, Newark, NJ 07102, USA}
\affiliation{NASA Ames Research Center, Moffett Field, CA 94035, USA}

\author[0000-0003-4627-8967]{Larisza Krista}
\affiliation{Cooperative Institute for Research in Environmental Sciences, University of Colorado, Boulder, CO 80309, USA}
\affiliation{National Centers for Environmental Information, National Oceanic and Atmospheric Administration, Boulder, CO 80305, USA}

\author{Sangwoo Lee}
\affiliation{Korean Space Weather Center, Jeju, Republic of Korea}

\author{Chris Lowder}
\affiliation{Southwest Research Institute, Boulder, Colorado, USA}

\author{Peter J.~MacNeice}
\affiliation{Heliophysics Science Division, 
NASA Goddard Space Flight Center, 
Greenbelt, MD 20771, USA}

\author{Astrid Veronig}
\affiliation{University of Graz, Institute of Physics, Graz, Austria}
\affiliation{Kanzelhöhe Observatory for Solar and Environmental Research, University of Graz, Graz, Austria}

\author{ISWAT Coronal Hole Boundary Working Team}

\begin{abstract}
Coronal holes are the observational manifestation of the solar magnetic field open to the heliosphere and are of pivotal importance for our understanding of the origin and acceleration of the solar wind. Observations from space missions such as the Solar Dynamics Observatory now allow us to study coronal holes in unprecedented detail. Instrumental effects and other factors, however, pose a challenge to automatically detect coronal holes in solar imagery. The science community addresses these challenges with different detection schemes. Until now, little attention has been paid to assessing the disagreement between these schemes. In this COSPAR ISWAT initiative, we present a comparison of nine automated detection schemes widely-applied in solar and space science. We study, specifically, a prevailing coronal hole observed by the Atmospheric Imaging Assembly instrument on 2018 May 30. Our results indicate that the choice of detection scheme has a significant effect on the location of the coronal hole boundary. Physical properties in coronal holes such as the area, mean intensity, and mean magnetic field strength vary by a factor of up to $4.5$ between the maximum and minimum values. We conclude that our findings are relevant for coronal hole research from the past decade, and are therefore of interest to the solar and space research community.

\end{abstract}

\keywords{Sun --- coronal holes  --- automated detection schemes --- ambient solar wind}

\section{Introduction}

\begin{figure*}
\begin{center}
\includegraphics[width=0.99\textwidth]{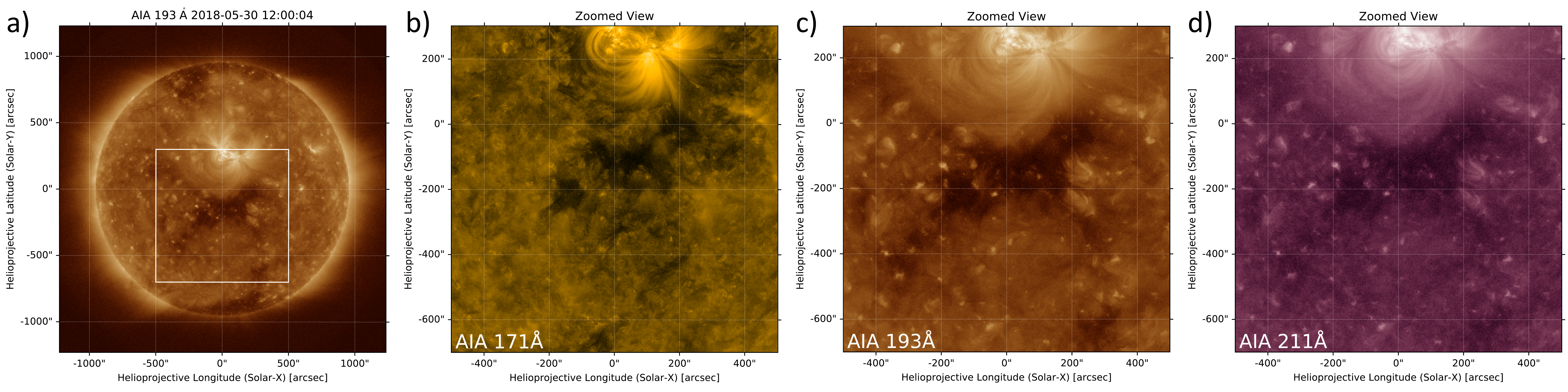}   
\end{center}
\caption{Example coronal hole from 2018~May~30 observed by the Solar Dynamics Observatory. (a) full-disk image of the Sun in the Fe~XII (19.3~nm; T~$\approx1.6$~MK) emission line; (b)--(d) coronal hole under scrutiny in the Fe~IX (17.1~nm; T~$\approx0.6$~MK), Fe~XII (19.3~nm; T~$\approx1.6$~MK), and Fe~XIV (21.1~nm; T~$\approx2.0$~MK) emission lines.\label{fig1}}
\end{figure*}

Coronal holes are an observational manifestation of open magnetic field lines emerging from the solar photosphere into interplanetary space. The evolving ambient solar wind is formed as coronal plasma escapes along these open field lines into space. Depending on the location and strength of the heating along the open field lines, several types of solar wind originate from coronal holes~\citep{mccomas07}. These types include highly Alfvénic slow wind emerging from coronal hole boundaries, and fast wind streams emerging from slowly diverging field lines deep inside low-latitude coronal holes and polar coronal hole extensions~\citep{zirker77,wang90}. At Earth, these fast solar wind streams can interact with the magnetosphere, causing geomagnetic disturbances~\citep{krieger73,tsurutani06}.

A defining feature of coronal holes is their reduced coronal emission. Because the open field guides coronal plasma into space, coronal holes are cooler and less dense than closed-field regions. They therefore emit less radiation than the adjacent coronal plasma. Ground and space-based instruments observe these reduced intensity regions at different wavelengths. The spectrum includes radio, near-infrared (particularly He~I 1083~nm), white light, EUV, and X-rays~\citep[see][]{newkirk67,munro72}. Due to these collective observations and advancements in instrumentation and remote sensing, coronal hole research has flourished over the past decades. This research has covered plasma and magnetic properties~\citep{zirker77, cranmer09}, temporal and spatial evolution, and the role played by coronal holes in modeling and predicting the ambient solar wind at Earth~\citep{wang90, riley01, arge03}. 

Automated coronal hole detection schemes are of broad interest to the community. Historically, He~I images have been used by observers to detect coronal holes by eye~\citep[][]{harvey02,mcintosh03}. The first automated coronal hole detection scheme using ground-based observations was developed by~\cite{henney05}. Later, automated schemes using a combination of ground and space-borne observations were proposed~\citep[see, for instance,][]{malanushenko05, detoma05, scholl08}. With approximately 70,000 images a day captured by the Solar Dynamics Observatory~\citep[SDO;][]{pesnell12} mission, the automated detection of coronal holes is an important diagnostic in solar and space science. Many unsolved questions in solar wind physics, such as the mechanism for solar wind acceleration and the origin of the slow solar wind, are intimately connected with coronal hole research~\citep{kilpua16, viall20}. 

But detecting coronal holes without human interaction is challenging~\citep{detoma05}. First, reduced coronal emission does not uniquely define coronal holes because filaments and regions of weak magnetic flux appear at a similar intensity; \citet{reiss15} showed that about 15\% of coronal holes detected with automated schemes are not coronal holes. Second, extracting coronal holes from photospheric magnetic field measurements is impossible~\citep{harvey79}; coronal holes are expected to be predominantly of one polarity but not all holes show this textbook behavior~\citep{hofmeister17, hofmeister19}. Third, the coronal hole appearance may vary greatly between different EUV and SXR filters sensitive to different plasma temperatures, which makes a definition of their boundaries difficult. Fourth, other important factors such as the noisy nature of EUV images, changes in viewing angle due to solar rotation \citep{wang17}, overshining of coronal hole regions due to large nearby coronal loops \citep{wang17}, systematic instrument effects, and the spatial and temporal evolution of coronal holes complicate their detection~\citep{caplan16}. As a consequence, coronal hole boundaries computed from automated schemes, which deal with these hindrances differently, are expected to show inherent discrepancies.

So a natural question arises: how large are the observational uncertainties of coronal hole boundaries in automated detection schemes? In 2019 the Coronal Hole Boundary Working Team\footnote{\url{https://iswat-cospar.org/S2-01}} in the COSPAR ISWAT initiative\footnote{\url{https://iswat-cospar.org}} was formed to address this question. In this letter we present our first results. We compare nine established detection schemes applied to an example coronal hole on 2018~May~30. We will show that the choice of scheme has a significant effect not only on the location of the coronal hole boundary but also on the inferred physical conditions inside the hole.

Our letter is outlined as follows. In Section~\ref{sec:event}, we present the example coronal hole on 2018~May~30. Section~\ref{sec:methods} describes the solar imagery and automated detection schemes, while Section~\ref{sec:results} presents the observational uncertainty of coronal holes in automated schemes. The discussion in Section~\ref{sec:discussion} concludes this letter and outlines future perspectives.

\begin{deluxetable*}{llllc}
\tablenum{1}
\tablecaption{Automated coronal hole detection schemes widely-applied in the community. \label{tab:schemes}}
\tablewidth{0pt}
\tablehead{
\colhead{Short} & \colhead{Research} & \nocolhead{}  & \colhead{Input} & \colhead{Online} \\
\colhead{Name} & \colhead{Institution} & \colhead{Reference} & \colhead{Waveband} & \colhead{Platform}}\decimalcolnumbers
\startdata
ASSA CH 		& Korean Space Weather Center      & \citet{hong12}     	    	& 19.3~nm	& \href{http://www.spaceweather.go.kr/assa}{Link}		\\
CHARM            & Trinity College Dublin; NOAA     					& \citet{krista09}             	& 19.3~nm, LOS~magnetogram 	& \href{https://github.com/lariszakrista/CHARM}{Link}                            \\
CHIMERA  	& Trinity College Dublin     		& \citet{garton18}   	     	& 17.1~nm, 19.3~nm, 21.1~nm      	& \href{https://github.com/TCDSolar/CHIMERA}{Link}                          \\
CHORTLE   	& Southwest Research Institute  	& \citet{lowder14}     		& 19.3~nm, LOS~magnetogram 	& \href{https://github.com/lowderchris/CHORTLE}{Link}     \\ 
CNN193           	& Moscow State University      		& \citet{illarionov18} 		& 19.3~nm	& \href{https://github.com/observethesun/coronal_holes}{Link}  \\
CHRONNOS           	& University of Graz      		& \citet{jarolim21} 		& 9.4~nm, 13.1~nm, 17.1~nm,	& -  \\
           	&       		&  		& 19.3~nm, 21.1~nm, 30.4~nm, 33.5~nm,	&   \\
           	           	&       		&  		& LOS~magnetogram	&   \\
SPoCA-CH             	& Royal Observatory of Belgium    	& \citet{delouille18}       	& 19.3~nm   	& \href{http://swhv.oma.be/user_manual/}{Link}                      \\ 
SYNCH            	& University of Oulu      			& \citet{hamada18}         	& 17.1~nm, 19.3~nm, 30.4~nm   	& -                             \\ 
TH35		& 		& -		      	& 19.3~nm   	& -                             \\ 
\enddata
\tablecomments{TH35 is a baseline reference scheme against which future studies can be easily compared against.}
\end{deluxetable*}

\section{A Case Study for 2018~May~30}\label{sec:event}
To quantify the uncertainty of coronal hole boundaries in automated detection schemes, we study an example low-latitude coronal hole on 2018~May~30. As shown in Figure~\ref{fig1}(a), the hole is located near the disk center, southward of the active region AR~12712. Persistent for several solar rotations, it was first observed around 2017~November~24. Initially, the hole was connected to a southern polar coronal hole with negative polarity as expected for solar cycle 24~\citep{lowder17}. After the example shown for 2018~May~30, the coronal hole was observable for five Carrington rotations before it disappeared.

As shown in the supplementary material, the coronal hole was associated with a high-speed stream in measurements by the Solar Wind Electron Proton Alpha Monitor~\citep[SWEPAM;][]{mccomas98} on-board the Advanced Composition Explorer~\citep[ACE;][]{stone98} spacecraft. Starting at around 2018~May~31~14:00~UT, the measurements show a gradual increase in bulk speed. On 2018~June~1 the solar wind speed peaked at around 700~km~s$^{-1}$. An increase in the magnetic field and density lead to a minor geomagnetic disturbance with a maximum $K_p$ of 5 and a minimum $Dst$ of $-38$~nT. 

\section{Methods}\label{sec:methods}
\subsection{Observational Data Preparation}
We use full-disk images in multiple EUV wavebands and photospheric magnetic field measurements from the Atmospheric Imaging Assembly~\citep[AIA;][]{lemen12} and the Helioseismic and Magnetic Imager~\citep[HMI;][]{scherrer12} instruments onboard the SDO spacecraft~\citep{pesnell12}. Due to the high contrast between coronal holes and the adjacent plasma, the AIA 19.3~nm waveband is most widely used. This filter is centered around the Fe~XII emission line at 19.3~nm and is dominated by the emission of plasma at around 1.6~MK. As illustrated in Figure~\ref{fig1}(b)--(d), some automated schemes also use lower temperature filters such as Fe~IX~17.1~nm and Fe~XIV~21.1~nm.

We downloaded all images from the SDO data archive as level 1.0 data, to which basic data calibration methods had already been applied. Next, we used the SolarSoft~\footnote{\url{https://sohowww.nascom.nasa.gov/solarsoft/}} procedure \verb|aia_prep.pro| to apply geometric corrections like centering the images, thereby removing shifts between the different AIA filters, correcting the roll-angle to align E-W and N-S to the x- and y-axes of the image, as well as scaling all images to 0.6~arcsec per pixel.

We use HMI measurements of the line-of-sight (LOS) component of the photospheric magnetic field. The LOS magnetogram (collected every 45 seconds) closest in time to the AIA image was selected for analysis. We processed the HMI magnetogram with \verb|aia_prep.pro| to apply the same geometric correction procedures. The HMI magnetogram, therefore, matches AIA down to less than a pixel. In this way, we map the coronal hole boundaries onto the magnetogram to retrieve photospheric magnetic field information. 

We distributed the level 1.5 processed images as FITS files in the original size of $4096 \times 4096$ pixels to the participating groups to detect the coronal hole boundaries with each scheme. By using the same dataset for all the detection schemes, we rule out discrepancies due to the preprocessing of the imagery. The exception was the ASSA algorithm, which requires preprocessed synoptic AIA images scaled down to $1024 \times 1024$ pixels in .jpg format.

\begin{figure*}
\includegraphics[width=0.99\textwidth]{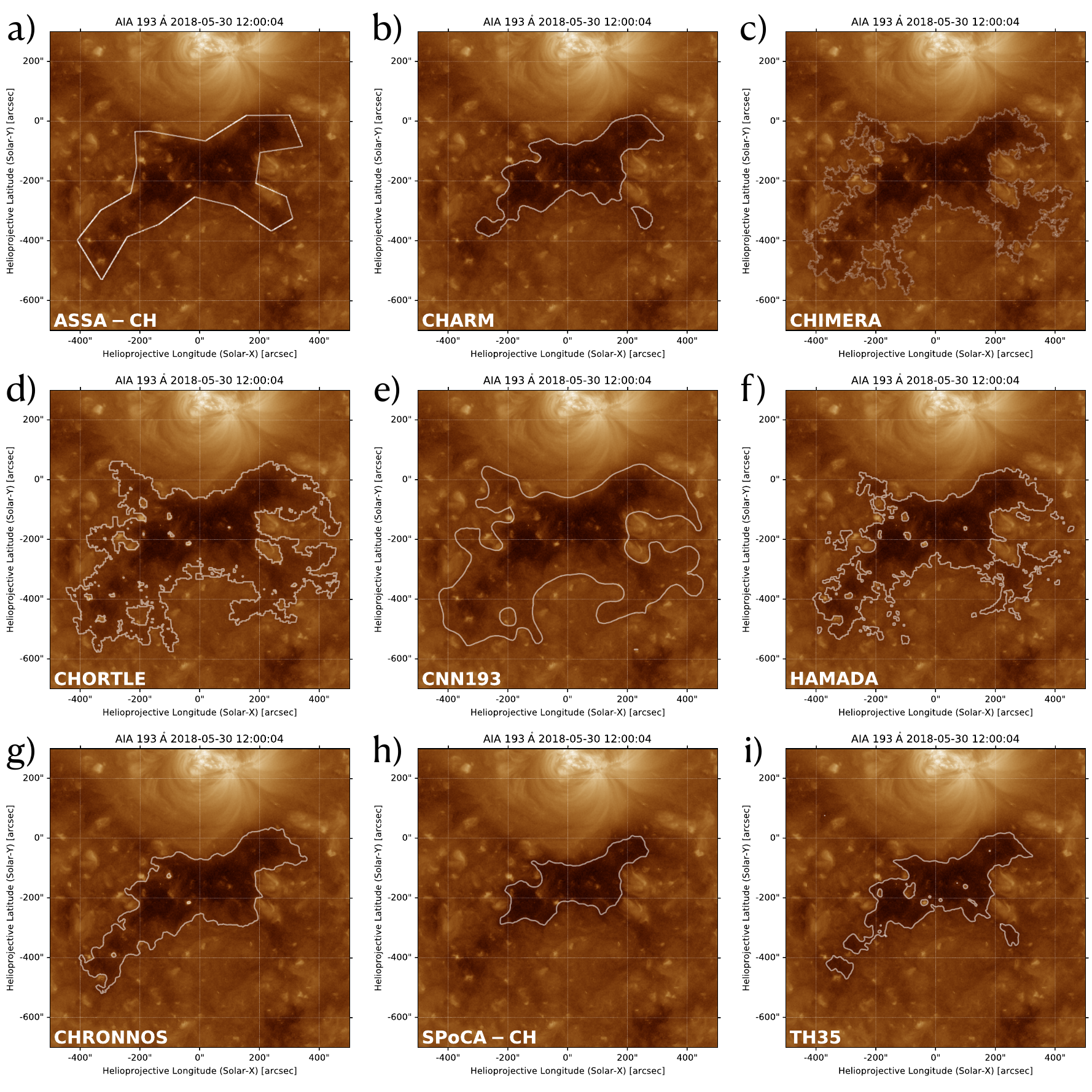}      
\caption{Differences of coronal hole boundary locations for the 2018~May~30 example due to the choice of the detection scheme. The background is an AIA 19.3~nm filtergram. (a) ASSA-CH; (b) CHARM; (c) CHIMERA; (d) CHORTLE; (e) CNN193; (f) HAMADA; (g) CHRONNOS; (h) SPoCA-CH; (i) TH35. 
\label{fig2}}
\end{figure*}

\subsection{Automated Detection Schemes}\label{sec:schemes}
We compare nine automated detection schemes widely applied in coronal hole research. Table~\ref{tab:schemes} lists these schemes along with their short name, institution, reference, input wavebands, and online platform. 

Although each scheme aims to classify each pixel in a solar coronal image as a coronal hole or background, the methods to do so are diverse. A common strategy is an intensity-based thresholding due to the high contrast of coronal holes compared to the rest of the coronal plasma, especially in the 19.3~nm waveband. The key is to find an intensity threshold that best separates the intensity distribution associated with coronal holes in an histogram. ASSA, for example, computes the intensity threshold equal to 45\% of the median intensity on the disk~\citep{hong12}. In contrast, SPoCA-CH relies on an iterative clustering algorithm called fuzzy C-means, which minimizes the variance in each cluster~\citep{verbeeck14}. The parameter setting used in the present study first determines the mode of the pixel intensity distribution, and then clusters in four classes the intensities smaller than this mode. The class with lowest intensity determines the coronal hole locations~\citep{delouille18}. These settings give different results from what is currently implemented in the SDO Event Detection System and the JHelioviewer.

\begin{figure*}
\includegraphics[width=0.99\textwidth]{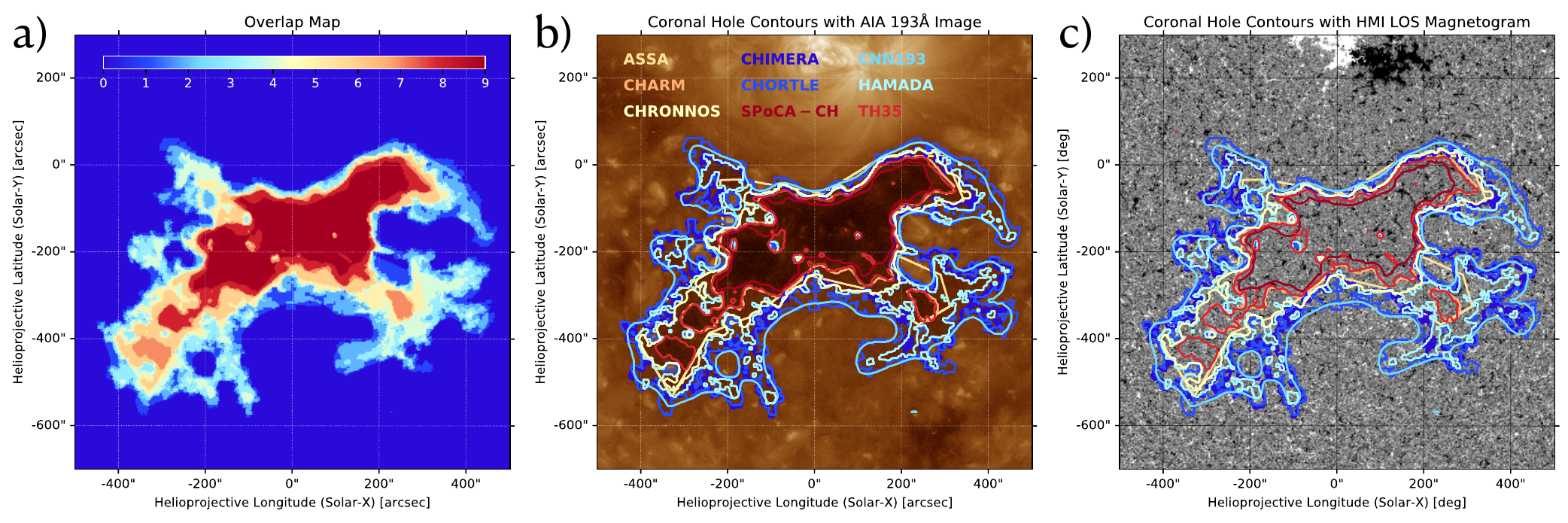}   
\caption{A comparison of the coronal hole maps from nine different automated detection schemes. (a) number of overlapping coronal hole detections; (b) coronal hole contours overlaid on an AIA 19.3~nm image; (c) the same contours overlaid on an HMI LOS magnetogram saturated at $\pm$~30~Gauss. 
\label{fig3}}
\end{figure*}

An alternative strategy uses thresholding by partitioning a full-disk image into sub-frames. This reduces the overlap between different features in an intensity histogram and the local minima are more clearly discernible. CHARM~\citep{krista09} and CHORTLE~\citep{lowder14} rely on this idea, searching for local minima in sub-frames of varying sizes. Measures such as the mean of the local minima define the global threshold. Furthermore, the intensity thresholding method in CHARM also relies on the overall intensity of the Sun, and hence the CH threshold detection adapts to the changing overall coronal emission over the solar activity cycle. 

The HAMADA scheme searches for the sub-frame that separates coronal holes most clearly. This search is done in three passbands (17.1~nm, 19.3~nm, 30.4~nm) in synoptic EUV maps, or alternatively in line-of-sight disk images. The logical conjunction of these three detections gives the final coronal hole map~\citep{hamada18}. Since coronal holes are often present in all three passbands, such an approach should avoid incorrectly segmented regions such as filaments. CHIMERA also uses three passbands but does not partition the solar image into sub-frames. Instead, the multi-thermal images are used to segment coronal holes by comparing the intensities across the three passbands. By computing a differentiation rule in the intensity space, CHIMERA creates coronal hole maps in all three wavebands, and the logical conjunction gives the final map~\citep{garton18}.

An original approach is taken in the CNN193 scheme. CNN193 uses a neural network to detect coronal holes in the 19.3~nm waveband. The network is trained on a dataset of solar images and semi-automatically created segmentation maps. Semi-automated means that the segmentation was done by an automated scheme but the results were supervised by an experienced observer. In this way, the neural network is a surrogate for coronal hole detections done by an observer~\citep{illarionov18}.

CHRONNOS also applies a neural network but uses all six EUV filtergrams of AIA and the HMI line-of-sight magnetogram simultaneously as input to the network. As a reference, it builds upon manually reviewed SPoCA-CH segmentation masks from~\citet[][]{delouille18}. With this approach, the identification of the coronal holes and their boundaries is based on the multi-channel EUV appearance and the underlying magnetic field. The network uses a progressively growing approach to include more spatial information, while also accounting for global relations in the full-disc observations~\citep[][]{jarolim21}.

To set a baseline against which future studies can easily be compared, we define a reference benchmark scheme called TH35. The intensity threshold is computed as 35\% of the medium intensity on the solar disk, with no further post-processing of the coronal hole maps. 

For an in-depth review of the automated schemes, we refer to the references in Table~\ref{tab:schemes}.

\subsection{Physical Properties in Coronal Holes} \label{sec:measures}
We study the physical properties inside the detected coronal holes using measures described in~\citet[][]{ko14}. For the EUV data we focus on the mean intensity in the AIA 19.3~nm waveband ($I_{193}$) which is the average intensity of all pixels inside the CH boundary given in data numbers (DNs) per second. 

In addition, we study several measures computed from HMI magnetograms, such as the signed ($B_{\text{LOS}}$) and unsigned ($|B_{\text{LOS}}|$) magnetic field strength in Gauss [G] averaged over the area outlined by the coronal hole boundary determined by each scheme. $B_{\text{LOS}}$ gives the net unbalanced field strength, which cancels the background noise that we assume is present in equal measure in both polarities. As such, it is a robust measure of the average field strength of the open field~\citep{abramenko09}. On the other hand, $|B_{\text{LOS}}|$ is the absolute value of the magnetic field strength that is related to the heating of the corona and the acceleration of the solar wind. 

Based on these two measures, the degree of unipolarity ($U$) is calculated using
\begin{equation}
U = \frac{\text{avg}(|B_{\text{LOS}}|) - |\text{avg}(B_{\text{LOS}})|}{\text{avg}(|B_{\text{LOS}}|)},
\end{equation}
where $U=0$ represents a pure unipolar field and $U=1$ represents a pure bipolar magnetic field~\citep[see][]{ko14}. It can be understood intuitively as a measure of how much the coronal hole differs from the expected textbook behavior.

Furthermore, we study the open magnetic flux defined as
\begin{equation}
\Phi = \sum_i^N B_{\text{LOS}}^{(i)} A^{(i)},
\end{equation}
where $N$ is the total number of coronal hole pixels, $B_{\text{LOS}}^{(i)}$ is magnetic field strength at each pixel, and $A^{(i)}$ is the corresponding pixel area. In this context, $\Phi$ represents the net flux through the coronal hole regions detected by the schemes in units of Maxwell [Mx].

\section{Results} \label{sec:results}

In Figure~\ref{fig2}, we present the coronal hole boundaries for the nine detection schemes on 2018~May~30, an example that is challenging for automated schemes. An initial visual assessment shows that the different schemes produce significantly different outcomes. These differences are minimal around the dark regions denoting the center of the coronal hole, and most prominent farthest from the center. Smaller differences also arise inside the coronal hole boundaries, most likely due to ephemeral regions inside the coronal hole~\citep[see, for instance,][]{wang20}.

While all methods are capable of detecting the coronal hole, its size and form vary considerably. And these differences are not only restricted to its shape; the physical properties of the coronal hole also show large differences. 

In Figure~\ref{fig3}(a), we quantify the uncertainties in the observed CH boundaries by assigning each pixel a number between 0 and 9, which reflects how many times the pixel was identified belonging to the coronal hole: e.g.~pixels with 9 have been identified by all nine detection schemes to belong to the coronal hole. In Figure~\ref{fig3} we also show the contours of the nine different CH boundaries over the AIA 19.3 nm image (b) and the HMI magnetogram (c). All schemes capture the darkest regions inside the boundary, but several also identify larger regions leading for instance to higher mean EUV emission.

\begin{figure*}
\includegraphics[width=0.99\textwidth]{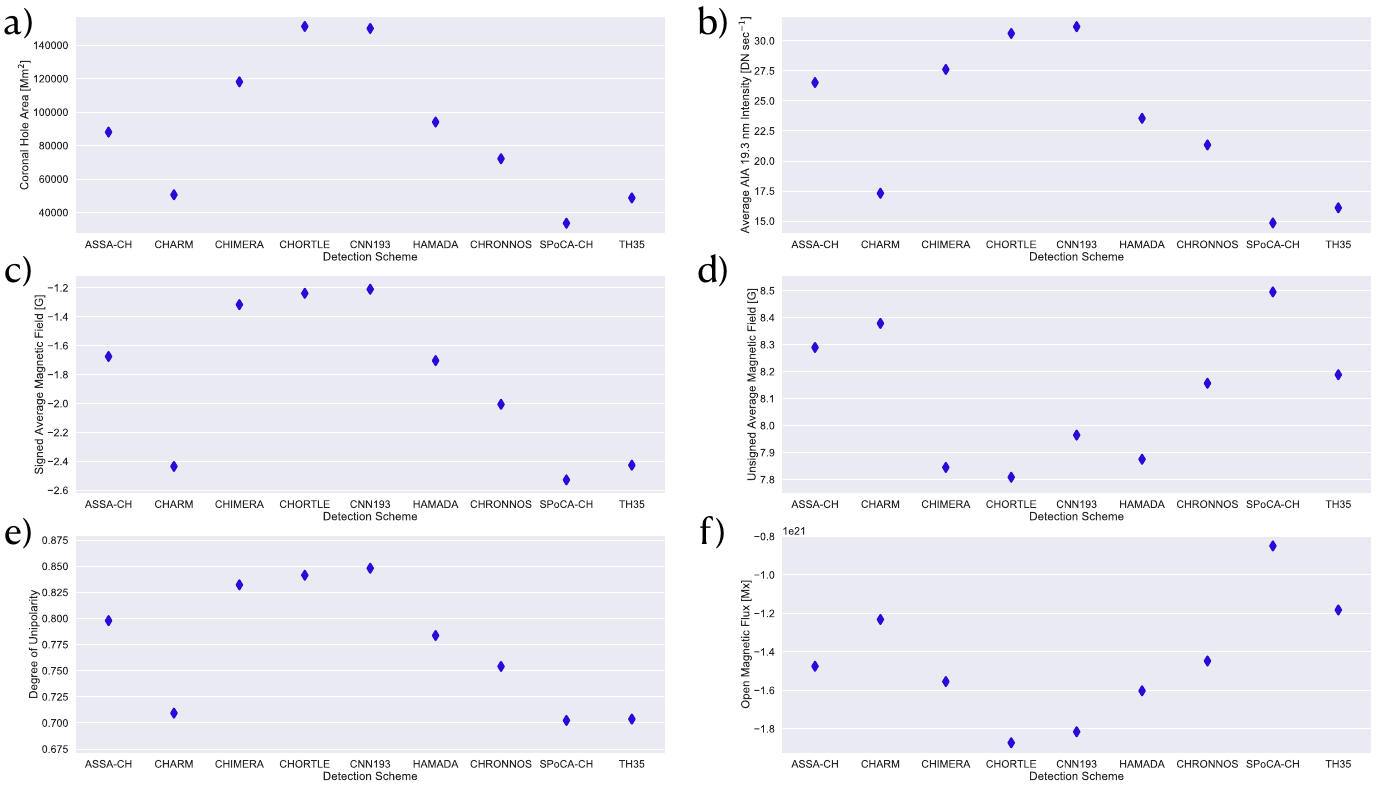}   
\caption{Range of physical properties inside coronal holes due the choice of the detection scheme. (a) coronal hole areas; (b) average AIA intensity in the 19.3~nm waveband; (c) signed average magnetic field strength; (d) unsigned average magnetic field strength; (e) degree of unipolarity; (f) open magnetic flux. 
\label{fig4}}
\end{figure*}

In Figure~\ref{fig4}, we show the coronal hole properties derived for the all detection schemes. Figure~\ref{fig4}(a) shows the coronal hole area, which ranges from $33.59 \times 10^3$~$\text{Mm}^2$ to $151.24 \times 10^3$~$\text{Mm}^2$. This means that the areas derived by the different schemes vary by a factor of $4.5$ between the minimum and maximum values. The mean AIA intensity in the 19.3~nm waveband (b) varies between 14.84~DN/sec and 31.18~DN/sec, with a factor of $2.1$ between the maximum and minimum value. Similarly, the signed average field strength (c) ranges from -1.21~G to -2.53~G, with a factor of $2.1$ between the maximum and minimum value. The unsigned average magnetic field strength (d) ranges between values of 7.85~G and 8.50~G, with a factor of $1.1$. The degree of unipolarity (e) ranges from 0.70 to 0.85, with a factor of $1.21$. Finally, the net open magnetic flux (f) ranges from $-8.49 \times 10^{20}$~Mx and $-1.87 \times 10^{21}$~Mx, with a factor between the maximum and minimum value of $2.2$. These differences in the geometry and physical properties indicate that the choice of the detection scheme plays a pivotal role in the analysis of coronal holes.

\section{Discussion} \label{sec:discussion}
The use of automated schemes for coronal hole detection is of critical importance for delineating the solar magnetic field that is open to the heliosphere. Although uncertainties are expected when automated schemes are applied, the question of how large these uncertainties are due to the choice of scheme has not yet been answered. By studying the coronal hole from  2018~May~30, we have shown that the choice of detection scheme has a large effect on the location of the coronal hole boundary. Moreover, physical properties in coronal holes vary by a factor of up to 4.5 between the maximum and minimum values. We will discuss the relevance of these results for three different research topics: (1) the physical properties and evolution of coronal holes and their associated fast solar wind streams, (2) their location and appearance throughout the solar activity cycle, and (3) their role as an observational diagnostic in coronal magnetic models.

1.~Our findings are most directly linked to the formation, evolution and decay of coronal holes and their relation to fast streams in the evolving ambient solar wind~\citep{wang10,detoma11,ko14,krista18}. The latter relates the physical properties of coronal holes to the conditions in Earth's space weather environment and is used in space weather prediction~\citep{nolte76,robbins06,vrsnak07,rotter12,reiss16,garton18}. As shown in this study, the choice of scheme can significantly affect the coronal hole boundary location, which has not been taken into consideration in most past studies. Such investigations would benefit from the uncertainties deduced by our comparison, which are valuable for the interpretation of their results. In this context, our findings also support recent efforts to construct error boundaries as an inherent data product in automated schemes, which have previously only been deduced from varying the parameters within a single method~\citep{heinemann19}. 

2.~Besides focusing on one coronal hole, much community effort is going into understanding the global distribution of coronal holes throughout the solar activity cycle. During solar minimum, coronal holes mostly reside in polar regions, while later in the cycle they appear at lower latitudes~\citep{lowder17,hewins20}. Tracking coronal holes and their associated open magnetic flux is a valuable diagnostic of the solar activity cycle~\citep{harvey02, wang09}. Taking into account that the computed coronal holes and the coronal hole properties can vary significantly when studied with different schemes shines a new light on these investigations. This is particularly relevant for the open magnetic flux from low-latitude coronal holes. 

3.~In the broader context, observationally derived coronal holes are an important test of global magnetic models of the corona~\citep[][]{mackay02,yeates10}. An open problem is that the modeled open magnetic flux systematically underestimates the observed open magnetic flux~\citep{linker17}. Recently,~\cite{wallace19} found that manually drawn coronal hole maps match coronal model solutions well, but automated detection schemes did not yield the same agreement~\citep{lowder14, lowder17}. Our deduced observational uncertainty complements the uncertainty estimates of models that use photospheric field measurements from different solar observatories or by using results from different flux transport models. Continuation of this study will lead to automated coronal hole maps with inherent error boundaries derived from the observations, which in future research can be compared with coronal hole maps computed from magnetic models, thus taking a leap towards solving this open problem. 

A pending question that arises is whether the derived uncertainties are observable only in some cases or represent a general trend. In the next step of our study, we will compare the results for a larger number of coronal holes. Furthermore, we will study the following influences on coronal boundaries in greater depth: (i) wavelength of the EUV images used in the detection scheme, (ii) position of the coronal hole on the solar disk, and (iii) phase in the solar cycle. 

Due to the broad application of our results in coronal hole research and related studies in solar and space science, we conclude that our results are valuable for a better understanding of past and future studies related to coronal holes in the community. 

To allow a comparison of future detection schemes with our findings, all the SDO data and related coronal hole maps are available online\footnote{\url{https://doi.org/10.6084/m9.figshare.13397261}}.

\newpage
\acknowledgments
The authors thank Maria Kuznetsova, Mario Bisi, Hermann Opgenoorth, and the cluster moderators for their commitment to the COSPAR ISWAT initiative, which supported this research effort. M.A.R.~thanks NASA's Community Coordinated Modeling Center for financial travel support. The authors acknowledge the following organizations and programs: M.A.R., C.M., and R.L.B.~acknowledge the Austrian Science Fund (FWF): J4160-N27, P31659, P31521; K.M.~acknowledges support by the NASA HGI program (\# 80HQTR19T0028) and the NASA cooperative agreement NNG11PL10A; A.V.~and R.J.~acknowledge the European Union’s Horizon 2020 research and innovation program under grant agreement No.~824135 (SOLARNET). The results of the CHRONNOS code have been achieved using the Vienna Scientific Cluster (VSC) and the Skoltech HPC cluster ARKUDA. E.I.~acknowledges the RSF grant 20-72-00106.

\bibliography{bib2020}{}

\begin{thebibliography}{}
\expandafter\ifx\csname natexlab\endcsname\relax\def\natexlab#1{#1}\fi

\bibitem[{{Abramenko} {et~al.}(2009){Abramenko}, {Yurchyshyn}, \&
  {Watanabe}}]{abramenko09}
{Abramenko}, V., {Yurchyshyn}, V., \& {Watanabe}, H. 2009, \solphys, 260, 43

\bibitem[{{Arge} {et~al.}(2003){Arge}, {Odstrcil}, {Pizzo}, \&
  {Mayer}}]{arge03}
{Arge}, C.~N., {Odstrcil}, D., {Pizzo}, V.~J., \& {Mayer}, L.~R. 2003, in
  American Institute of Physics Conference Series, Vol. 679, Solar Wind Ten,
  ed. M.~{Velli}, R.~{Bruno}, F.~{Malara}, \& B.~{Bucci}, 190--193

\bibitem[{{Caplan} {et~al.}(2016){Caplan}, {Downs}, \& {Linker}}]{caplan16}
{Caplan}, R.~M., {Downs}, C., \& {Linker}, J.~A. 2016, \apj, 823, 53

\bibitem[{{Cranmer}(2009)}]{cranmer09}
{Cranmer}, S.~R. 2009, Living Reviews in Solar Physics, 6, 3

\bibitem[{{de Toma}(2011)}]{detoma11}
{de Toma}, G. 2011, \solphys, 274, 195

\bibitem[{{Delouille} {et~al.}(2018){Delouille}, {Hofmeister}, {Reiss},
  {Mampaey}, {Temmer}, \& {Veronig}}]{delouille18}
{Delouille}, V., {Hofmeister}, S.~J., {Reiss}, M.~A., {et~al.} 2018, {''Chapter
  15 - Coronal Holes Detection Using Supervised Classification} (Elsevier),
  365--395

\bibitem[{{Garton} {et~al.}(2018){Garton}, {Gallagher}, \& {Murray}}]{garton18}
{Garton}, T.~M., {Gallagher}, P.~T., \& {Murray}, S.~A. 2018, Journal of Space
  Weather and Space Climate, 8, A02

\bibitem[{{Hamada} {et~al.}(2018){Hamada}, {Asikainen}, {Virtanen}, \&
  {Mursula}}]{hamada18}
{Hamada}, A., {Asikainen}, T., {Virtanen}, I., \& {Mursula}, K. 2018, \solphys,
  293, 71

\bibitem[{{Harvey} \& {Sheeley}(1979)}]{harvey79}
{Harvey}, J.~W., \& {Sheeley}, N.~R., J. 1979, \ssr, 23, 139

\bibitem[{{Harvey} \& {Recely}(2002)}]{harvey02}
{Harvey}, K.~L., \& {Recely}, F. 2002, \solphys, 211, 31

\bibitem[{{Heinemann} {et~al.}(2019){Heinemann}, {Temmer}, {Heinemann},
  {Dissauer}, {Samara}, {Jer{\v{c}}i{\'c}}, {Hofmeister}, \&
  {Veronig}}]{heinemann19}
{Heinemann}, S.~G., {Temmer}, M., {Heinemann}, N., {et~al.} 2019, \solphys,
  294, 144

\bibitem[{{Henney} \& {Harvey}(2005)}]{henney05}
{Henney}, C.~J., \& {Harvey}, J.~W. 2005, in Astronomical Society of the
  Pacific Conference Series, Vol. 346, Large-scale Structures and their Role in
  Solar Activity, ed. K.~{Sankarasubramanian}, M.~{Penn}, \& A.~{Pevtsov}, 261

\bibitem[{{Hewins} {et~al.}(2020){Hewins}, {Gibson}, {Webb}, {McFadden},
  {Kuchar}, {Emery}, \& {McIntosh}}]{hewins20}
{Hewins}, I.~M., {Gibson}, S.~E., {Webb}, D.~F., {et~al.} 2020, \solphys, 295,
  161

\bibitem[{{Hofmeister} {et~al.}(2019){Hofmeister}, {Utz}, {Heinemann},
  {Veronig}, \& {Temmer}}]{hofmeister19}
{Hofmeister}, S.~J., {Utz}, D., {Heinemann}, S.~G., {Veronig}, A., \& {Temmer},
  M. 2019, \aap, 629, A22

\bibitem[{{Hofmeister} {et~al.}(2017){Hofmeister}, {Veronig}, {Reiss},
  {Temmer}, {Vennerstrom}, {Vr{\v s}nak}, \& {Heber}}]{hofmeister17}
{Hofmeister}, S.~J., {Veronig}, A., {Reiss}, M.~A., {et~al.} 2017, \apj, 835,
  268

\bibitem[{{Hong} {et~al.}(2012){Hong}, {Lee}, {Oh}, {Kim}, {Lee}, {Kim}, {Lee},
  {Moon}, \& {Lee}}]{hong12}
{Hong}, S., {Lee}, S., {Oh}, S., {et~al.} 2012, in AGU Fall Meeting Abstracts,
  Vol. 2012, SH13A--2239

\bibitem[{{Illarionov} \& {Tlatov}(2018)}]{illarionov18}
{Illarionov}, E.~A., \& {Tlatov}, A.~G. 2018, \mnras, 481, 5014

\bibitem[{Jarolim {et~al.}(2021)Jarolim, Veronig, Hofmeister, Heinemann,
  Temmer, Podladchikova, \& Dissauer}]{jarolim21}
Jarolim, R., Veronig, A., Hofmeister, S., {et~al.} 2021, in preparation

\bibitem[{{Kilpua} {et~al.}(2016){Kilpua}, {Madjarska}, {Karna}, {Wiegelmann},
  {Farrugia}, {Yu}, \& {Andreeova}}]{kilpua16}
{Kilpua}, E.~K.~J., {Madjarska}, M.~S., {Karna}, N., {et~al.} 2016, \solphys,
  291, 2441

\bibitem[{{Ko} {et~al.}(2014){Ko}, {Muglach}, {Wang}, {Young}, \&
  {Lepri}}]{ko14}
{Ko}, Y.-K., {Muglach}, K., {Wang}, Y.-M., {Young}, P.~R., \& {Lepri}, S.~T.
  2014, \apj, 787, 121

\bibitem[{{Krieger} {et~al.}(1973){Krieger}, {Timothy}, \&
  {Roelof}}]{krieger73}
{Krieger}, A.~S., {Timothy}, A.~F., \& {Roelof}, E.~C. 1973, \solphys, 29, 505

\bibitem[{{Krista} \& {Gallagher}(2009)}]{krista09}
{Krista}, L.~D., \& {Gallagher}, P.~T. 2009, \solphys, 256, 87

\bibitem[{{Krista} {et~al.}(2018){Krista}, {McIntosh}, \& {Leamon}}]{krista18}
{Krista}, L.~D., {McIntosh}, S.~W., \& {Leamon}, R.~J. 2018, \aj, 155, 153

\bibitem[{{Lemen}~et al.(2012)}]{lemen12}
{Lemen}~et al., J.~R. 2012, \solphys, 275, 17

\bibitem[{{Linker} {et~al.}(2017){Linker}, {Caplan}, {Downs}, {Riley}, {Mikic},
  {Lionello}, {Henney}, {Arge}, {Liu}, {Derosa}, {Yeates}, \&
  {Owens}}]{linker17}
{Linker}, J.~A., {Caplan}, R.~M., {Downs}, C., {et~al.} 2017, \apj, 848, 70

\bibitem[{{Lowder} {et~al.}(2017){Lowder}, {Qiu}, \& {Leamon}}]{lowder17}
{Lowder}, C., {Qiu}, J., \& {Leamon}, R. 2017, \solphys, 292, 18

\bibitem[{{Lowder} {et~al.}(2014){Lowder}, {Qiu}, {Leamon}, \&
  {Liu}}]{lowder14}
{Lowder}, C., {Qiu}, J., {Leamon}, R., \& {Liu}, Y. 2014, \apj, 783, 142

\bibitem[{{Mackay} {et~al.}(2002){Mackay}, {Priest}, \& {Lockwood}}]{mackay02}
{Mackay}, D.~H., {Priest}, E.~R., \& {Lockwood}, M. 2002, \solphys, 209, 287

\bibitem[{{Malanushenko} \& {Jones}(2005)}]{malanushenko05}
{Malanushenko}, O.~V., \& {Jones}, H.~P. 2005, \solphys, 226, 3

\bibitem[{{McComas} {et~al.}(1998){McComas}, {Bame}, {Barker}, {Feldman},
  {Phillips}, {Riley}, \& {Griffee}}]{mccomas98}
{McComas}, D.~J., {Bame}, S.~J., {Barker}, P., {et~al.} 1998, \ssr, 86, 563

\bibitem[{{McComas} {et~al.}(2007){McComas}, {Velli}, {Lewis}, {Acton},
  {Balat-Pichelin}, {Bothmer}, {Dirling}, {Feldman}, {Gloeckler}, {Habbal},
  {Hassler}, {Mann}, {Matthaeus}, {McNutt}, {Mewaldt}, {Murphy}, {Ofman},
  {Sittler}, {Smith}, \& {Zurbuchen}}]{mccomas07}
{McComas}, D.~J., {Velli}, M., {Lewis}, W.~S., {et~al.} 2007, Reviews of
  Geophysics, 45, RG1004

\bibitem[{{McIntosh}(2003)}]{mcintosh03}
{McIntosh}, P.~S. 2003, in ESA Special Publication, Vol. 535, Solar Variability
  as an Input to the Earth's Environment, ed. A.~{Wilson}, 807--818

\bibitem[{{Munro} \& {Withbroe}(1972)}]{munro72}
{Munro}, R.~H., \& {Withbroe}, G.~L. 1972, \apj, 176, 511

\bibitem[{{Newkirk}(1967)}]{newkirk67}
{Newkirk}, Jr., G. 1967, \araa, 5, 213

\bibitem[{{Nolte} {et~al.}(1976){Nolte}, {Krieger}, {Timothy}, {Gold},
  {Roelof}, {Vaiana}, {Lazarus}, {Sullivan}, \& {McIntosh}}]{nolte76}
{Nolte}, J.~T., {Krieger}, A.~S., {Timothy}, A.~F., {et~al.} 1976, \solphys,
  46, 303

\bibitem[{Pesnell {et~al.}(2012)Pesnell, Thompson, \& Chamberlin}]{pesnell12}
Pesnell, W.~D., Thompson, B.~J., \& Chamberlin, P.~C. 2012, Solar Physics, 275,
  3

\bibitem[{{Reiss} {et~al.}(2015){Reiss}, {Hofmeister}, {De Visscher}, {Temmer},
  {Veronig}, {Delouille}, {Mampaey}, \& {Ahammer}}]{reiss15}
{Reiss}, M.~A., {Hofmeister}, S.~J., {De Visscher}, R., {et~al.} 2015, Journal
  of Space Weather and Space Climate, 5, A23

\bibitem[{Reiss {et~al.}(2016)Reiss, Temmer, Veronig, Nikolic, Vennerstrom,
  Schoengassner, \& Hofmeister}]{reiss16}
Reiss, M.~A., Temmer, M., Veronig, A.~M., {et~al.} 2016, Space Weather, 14,
  2016SW001390

\bibitem[{{Riley} {et~al.}(2001){Riley}, {Linker}, \& {Miki{\'c}}}]{riley01}
{Riley}, P., {Linker}, J.~A., \& {Miki{\'c}}, Z. 2001, \jgr, 106, 15889

\bibitem[{{Robbins} {et~al.}(2006){Robbins}, {Henney}, \& {Harvey}}]{robbins06}
{Robbins}, S., {Henney}, C.~J., \& {Harvey}, J.~W. 2006, \solphys, 233, 265

\bibitem[{{Rotter} {et~al.}(2012){Rotter}, {Veronig}, {Temmer}, \& {Vr{\v
  s}nak}}]{rotter12}
{Rotter}, T., {Veronig}, A.~M., {Temmer}, M., \& {Vr{\v s}nak}, B. 2012,
  \solphys, 281, 793

\bibitem[{{Scherrer} {et~al.}(2012){Scherrer}, {Schou}, {Bush}, {Kosovichev},
  {Bogart}, {Hoeksema}, {Liu}, {Duvall}, {Zhao}, {Title}, {Schrijver},
  {Tarbell}, \& {Tomczyk}}]{scherrer12}
{Scherrer}, P.~H., {Schou}, J., {Bush}, R.~I., {et~al.} 2012, \solphys, 275,
  207

\bibitem[{{Scholl} \& {Habbal}(2008)}]{scholl08}
{Scholl}, I.~F., \& {Habbal}, S.~R. 2008, \solphys, 248, 425

\bibitem[{{Stone} {et~al.}(1998){Stone}, {Frandsen}, {Mewaldt}, {Christian},
  {Margolies}, {Ormes}, \& {Snow}}]{stone98}
{Stone}, E.~C., {Frandsen}, A.~M., {Mewaldt}, R.~A., {et~al.} 1998, \ssr, 86, 1

\bibitem[{{Toma} \& {Arge}(2005)}]{detoma05}
{Toma}, G.~D., \& {Arge}, C.~N. 2005, in Astronomical Society of the Pacific
  Conference Series, Vol. 346, Large-scale Structures and their Role in Solar
  Activity, ed. K.~{Sankarasubramanian}, M.~{Penn}, \& A.~{Pevtsov}, 251

\bibitem[{{Tsurutani} {et~al.}(2006){Tsurutani}, {Gonzalez}, {Gonzalez},
  {Guarnieri}, {Gopalswamy}, {Grande}, {Kamide}, {Kasahara}, {Lu}, {Mann},
  {McPherron}, {Soraas}, \& {Vasyliunas}}]{tsurutani06}
{Tsurutani}, B.~T., {Gonzalez}, W.~D., {Gonzalez}, A.~L.~C., {et~al.} 2006,
  Journal of Geophysical Research (Space Physics), 111, A07S01

\bibitem[{{Verbeeck} {et~al.}(2014){Verbeeck}, {Delouille}, {Mampaey}, \& {De
  Visscher}}]{verbeeck14}
{Verbeeck}, C., {Delouille}, V., {Mampaey}, B., \& {De Visscher}, R. 2014,
  \aap, 561, A29

\bibitem[{{Viall} \& {Borovsky}(2020)}]{viall20}
{Viall}, N.~M., \& {Borovsky}, J.~E. 2020, Journal of Geophysical Research:
  Space Physics, doi:10.1029/2018JA026005

\bibitem[{{Vr{\v s}nak} {et~al.}(2007){Vr{\v s}nak}, {Temmer}, \&
  {Veronig}}]{vrsnak07}
{Vr{\v s}nak}, B., {Temmer}, M., \& {Veronig}, A.~M. 2007, \solphys, 240, 315

\bibitem[{{Wallace} {et~al.}(2019){Wallace}, {Arge}, {Pattichis},
  {Hock-Mysliwiec}, \& {Henney}}]{wallace19}
{Wallace}, S., {Arge}, C.~N., {Pattichis}, M., {Hock-Mysliwiec}, R.~A., \&
  {Henney}, C.~J. 2019, \solphys, 294, 19

\bibitem[{{Wang}(2009)}]{wang09}
{Wang}, Y.~M. 2009, \ssr, 144, 383

\bibitem[{{Wang}(2017)}]{wang17}
---. 2017, \apj, 841, 94

\bibitem[{{Wang}(2020)}]{wang20}
---. 2020, \apj, 904, 199

\bibitem[{{Wang} {et~al.}(2010){Wang}, {Robbrecht}, {Rouillard}, {Sheeley}, \&
  {Thernisien}}]{wang10}
{Wang}, Y.~M., {Robbrecht}, E., {Rouillard}, A.~P., {Sheeley}, N.~R., J., \&
  {Thernisien}, A.~F.~R. 2010, \apj, 715, 39

\bibitem[{{Wang} \& {Sheeley}(1990)}]{wang90}
{Wang}, Y.-M., \& {Sheeley}, Jr., N.~R. 1990, \apj, 355, 726

\bibitem[{{Yeates} {et~al.}(2010){Yeates}, {Mackay}, {van Ballegooijen}, \&
  {Constable}}]{yeates10}
{Yeates}, A.~R., {Mackay}, D.~H., {van Ballegooijen}, A.~A., \& {Constable},
  J.~A. 2010, Journal of Geophysical Research (Space Physics), 115, A09112

\bibitem[{{Zirker}(1977)}]{zirker77}
{Zirker}, J.~B. 1977, Reviews of Geophysics and Space Physics, 15, 257

\end{thebibliography}
\bibliographystyle{aasjournal}

\appendix
\section{Supplementary Material}
\begin{figure}[h!]
\includegraphics[width=0.99\textwidth]{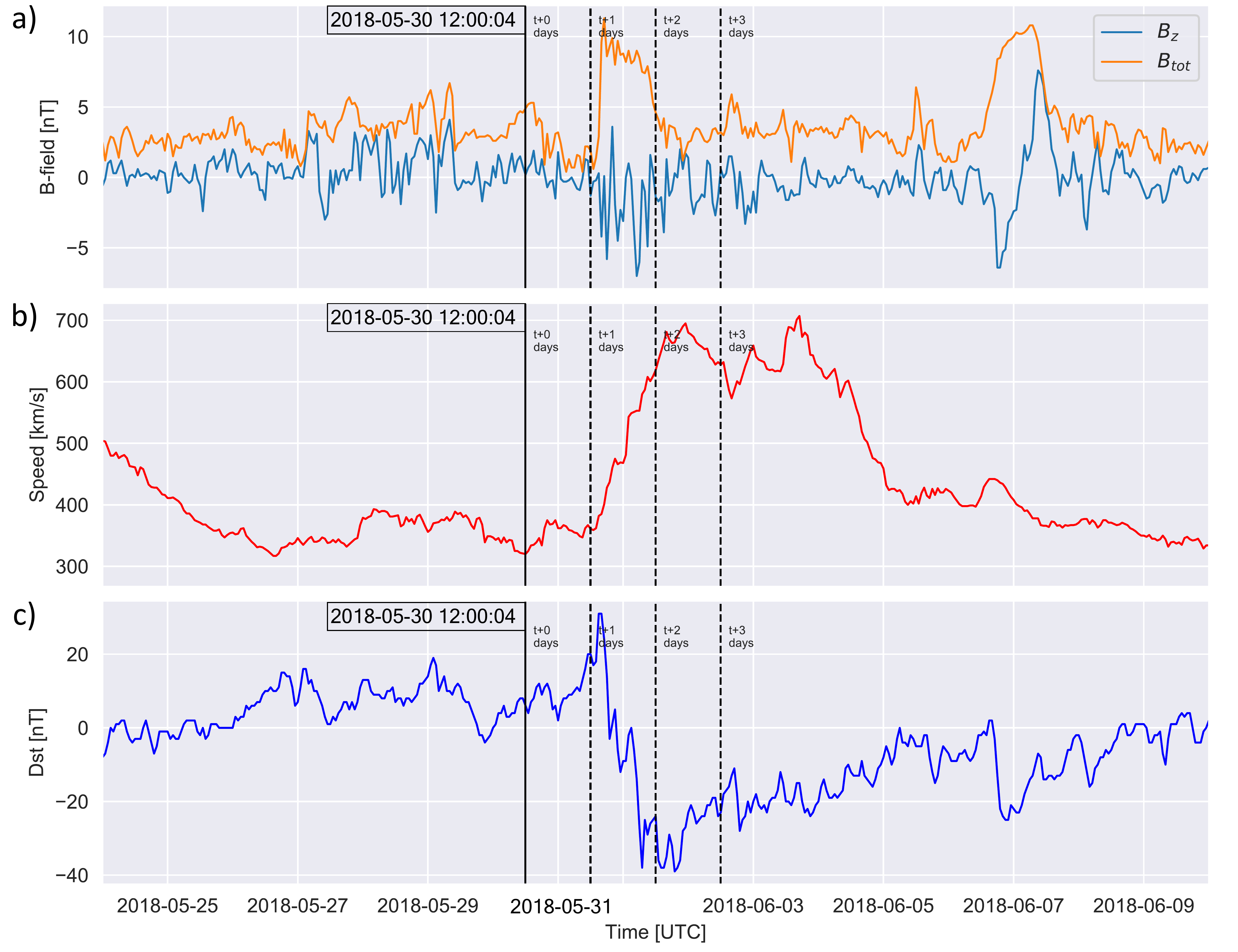} 
\caption{In-situ measurements at Earth of the solar wind associated with the coronal hole under scrutiny. (a) total magnetic field strength $B_{\text{tot}}$ and north-south pointing magnetic field component $B_{\text{z}}$; (b) solar wind bulk speed; (c) $Dst$ index as an indicator for geomagnetic activity. 
\label{supp1}}
\end{figure}
\end{document}